\documentclass[graybox,natbib,nosecnum]{svmult}
\bibpunct{(}{)}{;}{a}{}{,} 

\pdfoutput=1   

\usepackage{mathptmx}       
\usepackage{helvet}         
\usepackage{courier}        
\usepackage{type1cm}        

\usepackage{makeidx}         
\usepackage{graphicx}        
\usepackage{multicol}        
\usepackage[bottom]{footmisc}
\usepackage[normalem]{ulem}	
\usepackage{hyperref}  

\usepackage{soul}   

\newcommand*\vsini{\mbox{${\rm V}_{\rm rot}\sin i_{\star} $}}
\newcommand*\kms{\mbox{${\rm km\,s}^{-1} $}}
\newcommand*\vmac{\mbox{${\rm v}_{\rm mac} $}}

\newcommand{\hbindex}[1]{#1}  

\makeindex             

\begin{document}

\title*{Rotation of planet-harbouring stars}
\author{Pierre F. L. Maxted}
\institute{Pierre F. L. Maxted \at Astrophysics Group, Keele University,
Keele, Staffordshire, ST5 5BG, UK, \email{p.maxted@keele.ac.uk}}
%
%
\maketitle

\abstract{The rotation rate of a star has important implications for the
detectability, characterisation and stability of any planets that may be
orbiting it. This chapter gives a brief overview of stellar rotation before
describing the methods used to measure the rotation periods of planet host
stars, the factors affecting the evolution of a star's rotation rate, stellar
age estimates based on rotation, and an overview of the observed trends in the
rotation properties of stars with planets.}

\section{Introduction }

 Stars with convective envelopes, which includes most stars known to harbour
planets, show a variety of phenomena collectively known as \hbindex{magnetic
activity}. These include ultraviolet and X-ray emission, flares, and strong
emission lines at optical wavelength, particularly H$\alpha$ and the Ca\,II H
and K lines. All these phenomena are closely related to the rotation of the
star because it is the interaction of rotation and convection that drives the
generation of the complex surface magnetic field \citep{2004A&ARv..12...71G}.
This connection between rotation and magnetic activity has implications for
the detectability of planets orbiting solar-type stars, the environment in
which a planet exists, and can even lead to the destruction of a planet.
Detectable stellar rotation enables the alignment of a star's rotation axis
relative to the orbital plane of a planet to be measured, making it possible
to  study the dynamical history of a planetary system. 

\section{Overview of stellar rotation}
 Here I provide a brief overview of the observed rotational properties of
late-type main sequence stars, i.e. stars similar to most known planet host
stars.  A more general overview of stellar rotation can be found in
\cite{2013EAS....62..143B}. 

 It has been apparent since the earliest studies of stellar rotation that
late-type stars tend to rotate much more slowly than early-type stars
\citep{1930ApJ....72....1S}.  The rotation periods of main-sequence mid F-type
stars (T$_{\rm eff} \approx 6500$\,K, $M\approx 1.4M_{\odot}$) are typically a
few days, corresponding to equatorial rotation velocities $V_{\rm rot} \approx
20$\,--\,40\,\kms.  The mean value of  $V_{\rm rot}$ drops steady with spectral
type to a typical value of a few \kms\ for G-type main-sequence stars older
than about 1\,Gyr \citep{1982ApJ...261..259G}.  The rotation period
distribution becomes bimodal below mid-K spectral types with peaks near 20 and
40 days for mid M-type stars \citep{2014ApJS..211...24M, 2016ApJ...821...93N}.
The slow rotation of cool stars is a consequence of \hbindex{magnetic braking},
i.e., angular momentum loss through a magnetised stellar wind
\citep{1967ApJ...148..217W}. The age dependance of stellar rotation can be
studied using surveys of stellar rotation in open clusters. The behaviour at
very young ages is complex \citep{2013A&A...556A..36G}, but by the age of the
Hyades ($\sim 0.6$\,Gyr) there is a well-defined sequence of stars with
$P_{\rm rot}\approx 5$\,days at (B-V)=0.5 (F7) to $P_{\rm rot}\approx
12$\,days at (B-V)=1.2 (K5) \citep{1987ApJ...321..459R, 2016ApJ...822...47D}.
The rotation period then evolves approximately according to  $P_{\rm rot}
\propto t^{\gamma}$ with $\gamma\approx 0.5$ \citep{1972ApJ...171..565S}.
Exceptions to this general behaviour are discussed below in relation to
\hbindex{gyrochronology}, i.e., the use of stellar rotation to estimate the
age of a star. 

\section{Measuring stellar rotation}
 Methods used to study the rotation of stars in general are equally applicable to planet host stars. Additional information on the host star's rotation is available in transiting planet systems.

\runinhead{Rotational line broadening}
 
 The observed line profile of a rotating star is broader than the intrinsic
line profile emitted from each part of its surface because of the varying
Doppler shift across the visible hemisphere of the star.  The rotational
broadening is directly related to \vsini, where  $i_{\star}$ is the
inclination of the star's rotation axis to the line of sight. For a star with
rotation period $P_{\rm rot}$ and radius $R$, V$_{\rm rot} \approx 50.6
(R/R_{\odot})/(P/{\rm days})$\,\kms.  For a spherical star with no surface
\hbindex{differential rotation}, the shape of an intrinsically narrow stellar
absorption line with rest wavelength $\lambda_0$ is given by \begin{equation}
A(x) =  \frac{\frac{2}{\pi}\sqrt{1-x^2} +
\frac{\beta}{2}\left(1-x^2\right)}{1+\frac{2}{3}\beta}, \end{equation} where
$x=(\lambda/\lambda_0-1)/\left(\vsini/c \right)$ and $\beta$ is related to the
linear limb-darkening coefficient $u$ by   $\beta=u/(1-u)$
\citep{1955psmb.book.....U}. This rotational broadening function also assumes
that the limb darkening is the same at all wavelengths and that the intrinsic
line profile is constant over the visible surface of the star, and neglects
gravity darkening. These assumptions work well for the approximate \vsini\
range 3\,--\,30\,\kms, which covers the rotation rates for many planet host
stars.
  
 Equation (1) can be used to estimate the value of \vsini\ for a star from the
width of the absorption lines in its spectrum. This can be done directly from
the full-width at half-minimum (FWHM) of individual lines in a high-resolution
spectrum if the signal-to-noise is high enough (S/N $\approx 100$ or more).
The position of the first zero in the Fourier transform of the line profile
can be used to estimate \vsini\ \citep{1995ApJ...439..860C}. If the position
of the second zero of the Fourier transform can be measured then the
differential rotation of the star can be measured \citep{2002A&A...384..155R}.
Using individual lines to measure \vsini\ becomes difficult if isolated lines
cannot be identified or the signal-to-noise is not very high. In this case,
the observed spectrum can be compared to synthetic spectra that have been
convolved with the rotational broadening profile $A(x)$ to obtain an estimate
of \vsini. The observed spectra of slowly-rotating stars with a similar
spectral type to the target star can also be used as templates in this
approach. More often, these rotationally-broadened spectra are used to create
a calibration between \vsini\ and some global measure of the average line
width in the spectrum that can be easily calculated for large numbers of
stars, e.g., the FWHM of the \hbindex{cross-correlation function}
\citep{2010A&A...517A..88W}.

\begin{figure}
\includegraphics[scale=.65]{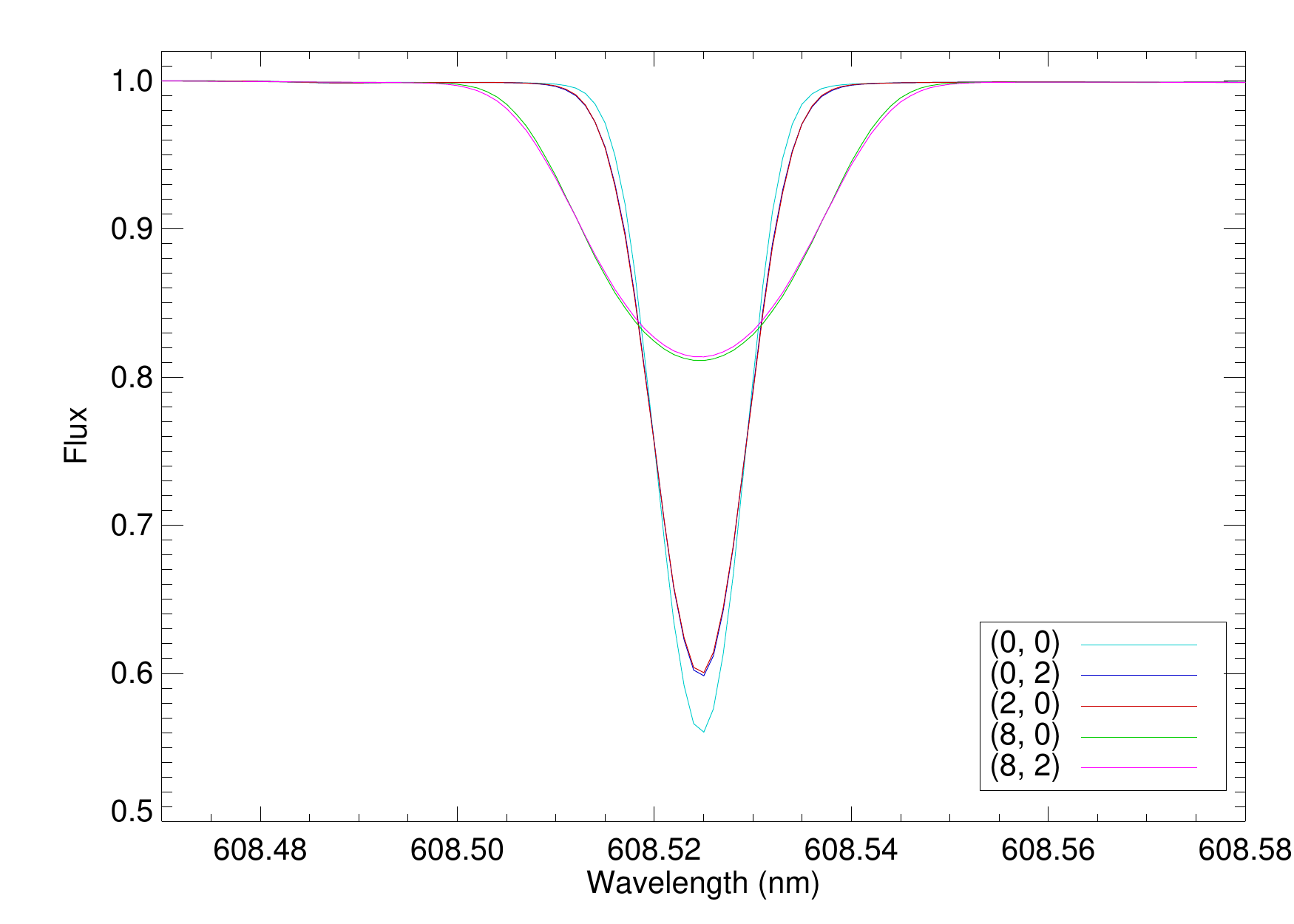}
\caption{Synthetic spectra in the region of an isolated Fe\,I line for a
Sun-like star. The values of (\vsini, \vmac) in km\,s$^{-1}$ are noted in the
legend. }
\label{fig:1}       
\end{figure}

 Estimating \vsini\ from using line broadening becomes difficult if the
rotational line broadening is comparable to the intrinsic line width. The
absorption lines in a stellar spectrum are broadened by various processes
acting at microscopic scales in the line forming region, e.g., thermal line
broadening and \hbindex{Zeeman splitting} due to magnetic fields. The
absorption lines are also broadened by turbulence in the photospheres of stars
with convective atmospheres. Convection can be simulated using 3-dimensional
radiation hydrodynamic models, but this is computationally expensive. For the
1-dimensional model atmospheres normally used for spectral analysis the effect
of turbulence on large scales (\hbindex{macroturbulence})  is approximated by
convolving the spectrum with a Gaussian profile of width \vmac.  Various
calibrations for the parameter \vmac\  have been published. These typically
give a value of  $\vmac\approx2$\,--\,4\,km\,s$^{-1}$ for the Sun with \vmac\
increasing with effective temperature up to $\vmac\approx6$\,km\,s$^{-1}$ at
T$_{\rm eff} =6500$\,K. Fig.~\ref{fig:1} shows the effect of rotational
broadening on synthetic spectra for a Sun-like star assuming
$\vmac=2$\,km\,s$^{-1}$ or $\vmac=4$\,km\,s$^{-1}$. This illustrates how
uncertainties in the intrinsic line profile of the star can lead to systematic
errors 1\,km\,s$^{-1}$ or more in the estimates of \vsini\ if this quantity is
a few \,km\,s$^{-1}$ or less. Similar problems can occur if \vsini\ is less
than the resolution of the spectrograph used to obtain the spectrum. Surface
differential rotation can also lead to systematic errors in estimating \vsini\
because the line profiles are the result of the integrated flux from the
visible stellar surface, not just the equator. For example,
\cite{1996A&AS..118..595V} find that $\vsini \approx 1.6$\,\kms\ provides the
best match to the width of the lines in solar spectrum neglecting differential
rotation,  significantly lower than the true value $\vsini= 2.0$\,\kms.

 The precision of a radial velocity measurement from one weak absorption line
can be estimated from $\sigma_v \approx {\rm FWHM}^{3/2}/(W\sqrt{I_0})$, where
$W$ is the \hbindex{equivalent width} of the line and $I_0$ is the
signal-to-noise ratio \citep{2015PASP..127.1240B}. There is also additional
noise in radial velocity measurements for cool stars (\hbindex{jitter}) due to
turbulence in their atmospheres and \hbindex{star spots}, particularly for
magnetically active stars \citep{2005PASP..117..657W}. This makes it easier to
find planets around slowly rotating stars that have narrower spectral lines
and are less magnetically active.

\begin{figure}
\includegraphics[scale=.65]{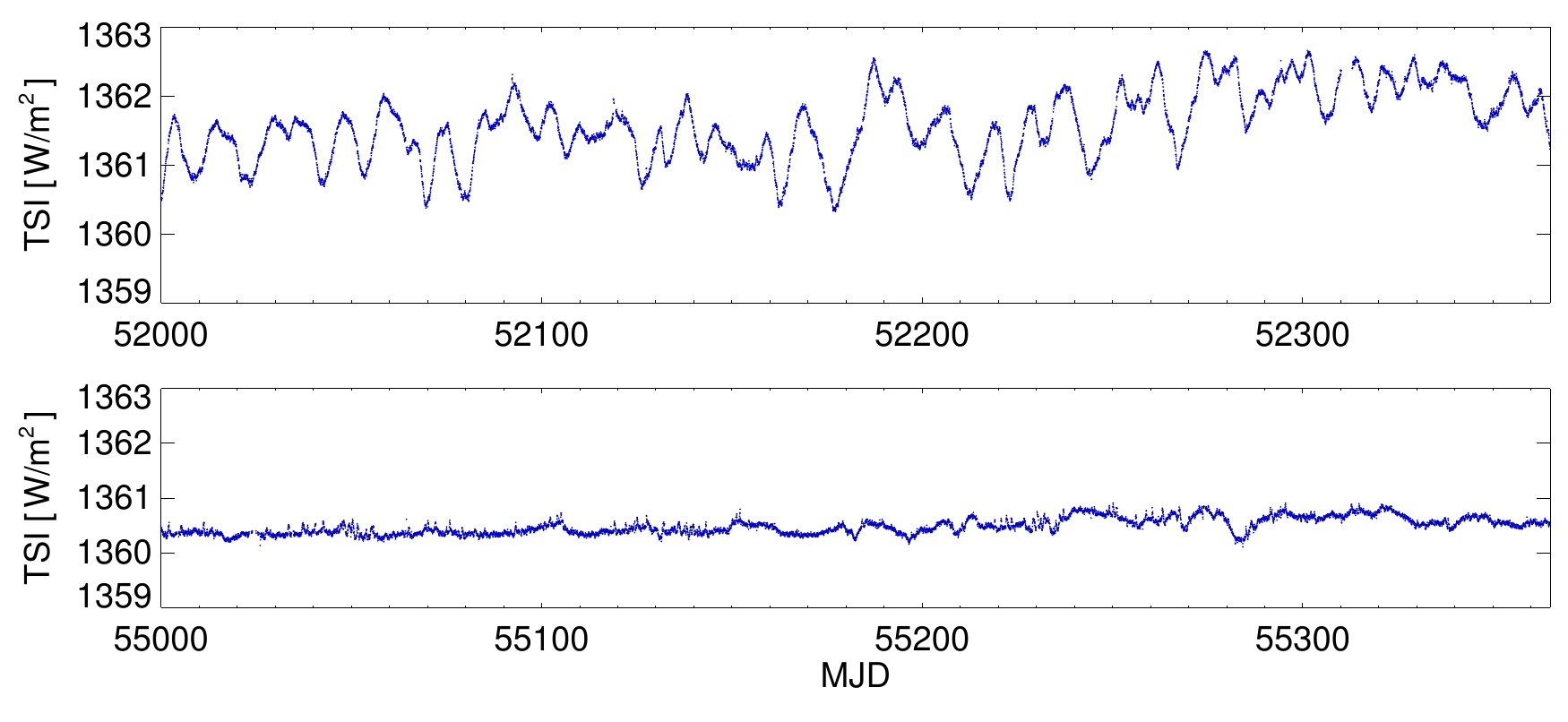}
\caption{Total solar irradiance (TSI) during solar maximum (upper panel) 
and solar minimum (lower panel)  \citep{VIRGO_TSI}. MJD is the modified Julian
Date.  
}
\label{fig:2}       
\end{figure}

\runinhead{Light curve  modulation}
 Total solar irradiance (TSI) varies quasi-periodically with a timescale
$\approx 26$\,days,  particularly during solar maximum (Fig.~\ref{fig:2}).
This is mostly due to the changing visibility of long-lived sunspots as the
Sun rotates. Similar variability is seen in the optical light curves of
solar-type stars and so the timescale of these variations is interpreted as
the rotation periods of these stars at the latitudes of their magnetically
active regions. The timescale of the variations is normally measured using
either Fourier methods \citep{2011PASP..123..547M}, from the auto-correlation
function \citep{2013MNRAS.432.1203M}, or using wavelet decomposition
\citep{2014A&A...572A..34G}. The rotation period derived from a light curve
with limited coverage can be ambiguous by a factor of two if there are two
large active regions on opposite hemispheres of the star. In general, there is
good agreement between the rotation periods derived from the modulation of the
flux and those obtained from the spectroscopic \vsini\ for stars where an
estimate of the radius is available.
 
\runinhead{Spot crossing events} If a planet crosses a dark spot on the
stellar surface during a transit there will be an anomalous ``bump'' in the
light curve.  With high quality photometry it may be possible to detect
consecutive spot crossing events and so infer the rotation period of the star
\citep{2008ApJ...683L.179S}. This method also provides  useful constraints on
the angle between the orbital axis of the planet and the rotation axis of the
star \citep{2013MNRAS.428.3671T, 2014ApJ...788....1B}.

\runinhead{Asteroseismology} 
\hbindex{Asteroseismology} uses frequency analysis of stellar light curves to
infer the interior structure, composition and other fundamental parameters of
pulsating stars. Solar-type stars show variations in brightness with periods
$\approx$ 5 minutes due to stochastic excitation of oscillation modes driven
by convection. These variations are small and have a complex spectrum so
extensive application of asteroseismology to solar-type stars is only
possible using very high-quality photometry from instruments such as
\hbindex{CoRoT} \citep{2009A&A...506..411A} and \hbindex{Kepler}
\citep{2010Sci...327..977B}. The frequency of the mode with radial order $n$
and  angular degree $\ell$ is split into  $(2\ell + 1)$ components according
to the azimuthal order $m$. For slowly-rotating Sun-like stars this splitting
is given by \begin{equation} \nu_{n,l,m} \approx \nu_{n,l} + m \Omega/2\pi,
\end{equation} where $\Omega$ is the average angular velocity in the outer
layers of the star \citep{2013PNAS..11013267G}. In additional, the amplitudes
of modes with different values of $m$ can be used to infer the inclination of
the rotation axis to the line of sight.  Asteroseismology can also be used to
estimate the mass and age of a star by comparing its pulsation frequencies and
other observed properties to the predictions of stellar evolutionary models.

\runinhead{\hbindex{Rossiter-McLaughlin effect}} 
 The transit of a star by a planet distorts the stellar absorption line
profiles. With very high quality spectra it is possible to observe this
distortion directly \citep{2016A&A...588A.127C}, but it is more common to
measure it indirectly using the mean Doppler shift of the stellar spectrum.
This apparent Doppler shift shows an anomalous variation during the transit
known as the Rossiter-McLaughlin (R-M) effect. The shape and amplitude of the
R-M effect for a star of radius $R_{\star}$ orbited by a planet of radius
$R_p$ with orbital inclination $i_p$ and semi-major axis $a$ depend mostly on
\vsini, the impact parameter $b=a\cos(i_p)/R_{\star}$, $k=R_p/R_{\star}$ and
the sky-projected angle between the stellar rotation axis and the orbital
angular momentum vector, $\lambda$. The line profile is asymmetric so there is
also a dependance on the method used to measure the apparent Doppler shift
\citep{2011ApJ...742...69H, 2005ApJ...622.1118O,2013A&A...550A..53B}. Doppler
tomography can be used to detect the distortion to the line profile directly
from an average line profile. This method works particularly well for rapidly
rotating stars such as the A-type planet host WASP-33, and may be the only
practical way to confirm the existence of a planetary companion to such stars
\citep{2010MNRAS.407..507C}.  

 Observations of the R-M effect generally focus on the derivation of $\lambda$
for transiting exoplanets, but can also provide useful constraints on \vsini.
The model degeneracies between \vsini\ and the other model parameters are
complex and can lead to implausible solutions for the least-squares fit to the
R-M effect, particularly when the detection of the effect is marginal. The
value of  \vsini\  derived from the line broadening can be imposed as a
constraint for the analysis in these cases \citep{2010A&A...524A..25T,
2012ApJ...757...18A}.

\section{Rotational evolution}

 The distortion of a star of mass $M_1$ and radius $R_1$ by a planet of mass
$M_2$ at a distance $d$ is approximately $\varepsilon = (M_2/M_1)(R_1/d)^3$,
so \hbindex{tides} are of greatest relevance to stars with massive
exoplanets in close orbits, i.e., hot Jupiter systems. This distortion varies
throughout the planet's orbit unless the star's rotation period equals the
planet's orbital period (synchronous rotation), the planet's orbit is
circular,  and the orbit is aligned with the star's rotation axis (zero
\hbindex{obliquity}). In general, the varying distortion of the star leads to
dissipation of energy in the star  and the exchange of angular momentum
between the rotation of a star and the orbit of the planet
\citep{2014ARA&A..52..171O}. For most hot Jupiter systems the angular momentum
is lost from the orbit of the planet resulting in the ``spin-up'' of the host
star and the eventual destruction of the planet \citep{2009ApJ...692L...9L,
2010ApJ...725.1995M}. The lifetimes of hot Jupiters are directly related to
the modified tidal quality factor of the star, $Q^{\prime}_{\star}$.
$Q^{\prime}_{\star}$ depends strongly on the properties of the star, and also
depends on the frequency and amplitude of the tidal distortion itself, i.e.,
tidal dissipation is a non-linear problem. As a result, it is very difficult
to predict an accurate value of $Q^{\prime}_{\star}$ for a given planetary
system. \citet{2009ApJ...692L...9L} adopt a value $Q^{\prime}_{\star} = 10^6$
to derive lifetimes of about 1\,Gyr for a typical hot Jupiter system, but
stress that this parameter could be anywhere in the range $10^5 <
Q^{\prime}_{\star} < 10^{10}$.
 
 The spin-up of a star with a convective envelope by a hot Jupiter companion
will lead to increased magnetic activity and, perhaps, an increase in
magnetic braking. This may lead to a long-lived pseudo-stable equilibrium
state for the star in which there is a balance between the angular momentum
loss by  magnetic braking and spin-up by a hot Jupiter companion
\citep{2004ApJ...610..464D, 2015A&A...574A..39D}. Hot Jupiters can also modify
the global structure of the stellar corona and the stellar wind, reducing the
efficiency of magnetic stellar wind braking \citep{2010ApJ...723L..64C}, i.e.,
hot Jupiters may prevent spin-down in some stars, as well as causing spin-up
due to tidal dissipation.

 The spin-up of a Sun-like star by a companion of 1 Jupiter mass is expected
to be negligible if their initial separation is larger than
$0.04$\,--\,0.08\,AU \citep{2015ApJ...807...78F}, i.e., initial periods longer
than 3\,--\,8 days, where the uncertainty in this estimate is dominated by the
assumed value of $Q^{\prime}_{\star}$. Tides are much weaker for stars with
smaller radii so the spin-up of  low mass stars (M-dwarfs) by planetary
companions is expected to be negligible if the planet survives the initial
contraction onto the main sequence \citep{2012A&A...544A.124B}.
For more massive stars, tidal dissipation and magnetic braking can lead to the
planet falling into the star during its main-sequence life time. Tidal
dissipation also plays a role in the engulfment of planets by stars as they
evolve to become red giants \citep{2016A&A...591A..45P}.  These merger events
may be directly observable \citep{2012MNRAS.425.2778M} and can alter the
surface composition of the star \citep{1999MNRAS.308.1133S}, and will almost
certainly result in  the spin-up of the star due to the accretion of the
planet's orbital angular momentum.

\section{Gyrochronology}
 The observation that older stars tend to rotate more slowly than younger
stars provides a means to estimate the age of a star based on its rotation
period, a technique known as gyrochronology. \citet{2010ApJ...722..222B}
recommends the following equation to estimate the age ($t$) for a star with
measured rotation period $P$: 
\begin{equation}
\label{gyro-eqn}
t = \frac{\tau}{k_C}\ln \left(\frac{P}{P_0}\right) +
\frac{k_I}{2\tau}\left(P^2-P_0^2\right),
\end{equation}
where $P_0 \approx1.1$\,d is the rotation period on the zero-age main sequence
(ZAMS) and $\tau$ is the \hbindex{convective turnover timescale}. The
constants $k_C$ and $k_I$ have the values $k_C =0.646$\,d/Myr and $k_I =
452$\,Myr/d when $\tau$ is taken from Table 1 of \citet{2010ApJ...721..675B}.
These values are chosen to match the rotational behaviour of stars in young
open clusters (t$\approx 150$\,Myr) and the rotation period of the Sun at its
current age. Magnetic activity does not increase with rotation rate above some
limit for stars with very rapid rotation.  The first term on the right hand
side of equation (\ref{gyro-eqn}) accounts for this saturation effect in young
stars. The second term is the Skumanich-like relation, $P_{\rm rot} \propto
t^{0.5}$ that dominates for older stars. This formulation provides a better fit
to the rotational behaviour of young stars than formulae that use only a
Skumanich-like relation. 

 Until recently, the calibration  of gyrochronology was based only on stars
in open clusters younger than 1\,Gyr plus the Sun. The \hbindex{Kepler
mission} has made it possible to test gyrochronology for stars in open
clusters with ages similar to the Sun (Fig.~3). Applying equation
(\ref{gyro-eqn}) to stars observed using Kepler in NGC~6819 and M67 gives
predicted ages of 2.5\,Gyr with a standard deviation of 0.25\,Gyr and 4.2\,Gyr
with a standard deviation of 0.7\,Gyr, respectively. These estimates are
consistent with other age estimates for stars in these clusters.

\begin{figure}
\includegraphics[scale=.65]{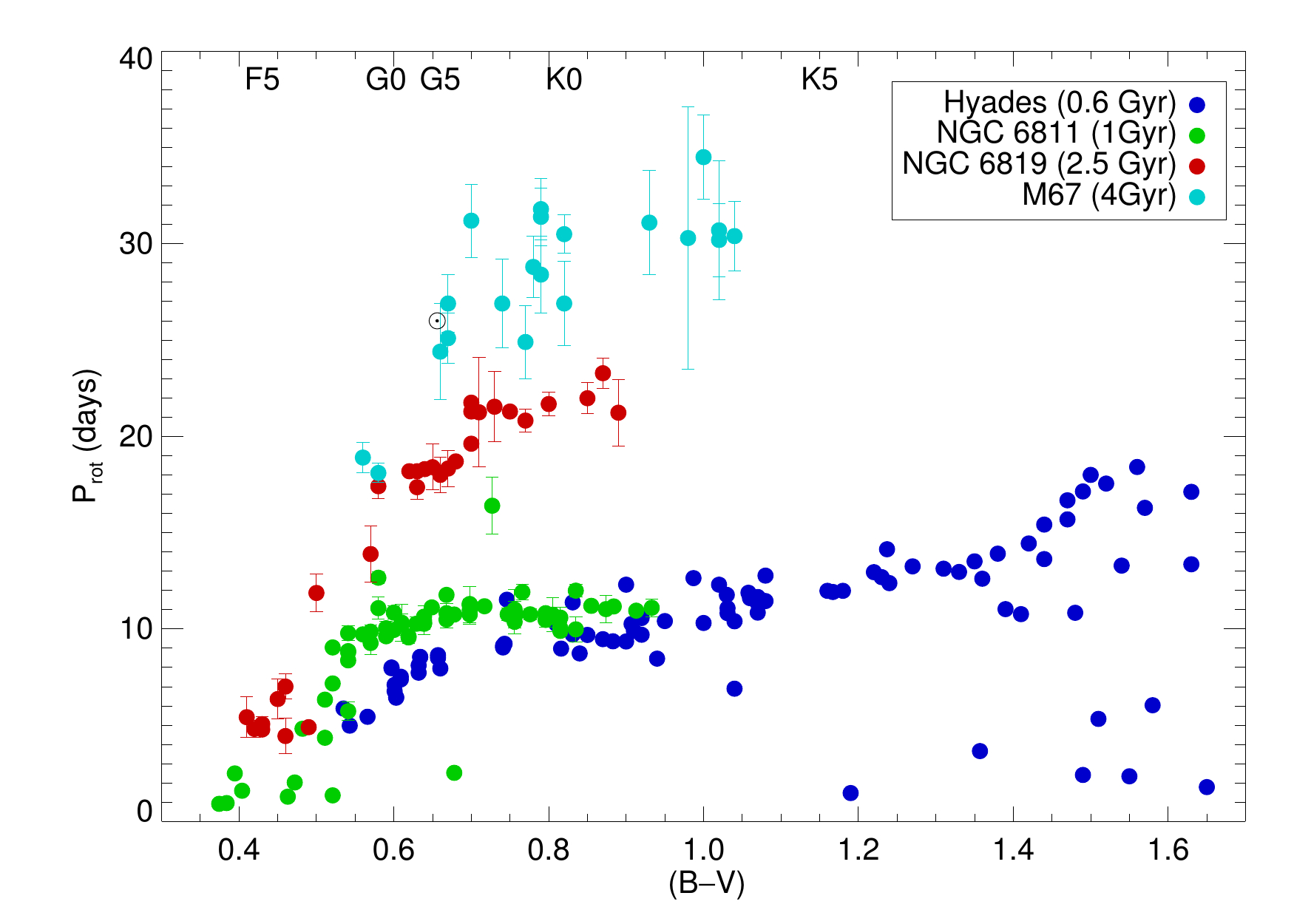}
\caption{Rotation periods of stars in selected open clusters and the Sun
($\odot$) as a function of (B$-$V) photometric colour
\citep{2016ApJ...823...16B, 2015Natur.517..589M, 1987ApJ...321..459R,
1995ApJ...452..332R, 2011ApJ...733L...9M}. 
}
\label{fig:3}       
\end{figure}

 Gyrochronology is only applicable to solar-type dwarf stars that spin-down
due to magnetic braking. It is not expected to give reliable age estimates
for the following types of star.
\begin{description}
\item[Stars in close binaries.]{Strong tidal interaction forces stars to
rotate synchronously with the orbital period in binary systems with orbital
periods $\approx  10$\,d or less. \citep{2010A&ARv..18...67T}. The question of
whether tidal dissipation affects gyrochronological ages for planet host stars
is discussed below.}
\item[Evolved stars.]{The internal structure of a star and its radius both
evolve rapidly when it reaches the main-sequence turn off.
\citet{2016AN....337..810B} recommend that gyrochronology only be applied to
stars with surface gravity  $\log g \ge 4.3$ (where the units for  $g$ are
cm\,s$^{-2}$).}
\item[Metal-poor and metal-rich stars]{Gyrochronology has thus far only been
calibrated using open cluster data for stars close to solar metallicity.
Magnetic braking may be different for stars with non-solar metallicity. }
\item[Very young stars]{ There is considerable scatter in the rotation periods
of stars on the ZAMS, from $P_0 = 0.2$ to 4.3\, days. Since magnetic braking
is more efficient for more rapidly rotating stars, this initial wide spread in
rotation periods converges with time to about 25\% by the age of the Hyades.
This behaviour can be modelled approximately using equation (\ref{gyro-eqn})
by varying $P_0$, but is not included at all in prescriptions for
gyrochronology that use only a Skumanich-like spin-down law. In either case,
the gyrochronological ages for young stars can be very uncertain.} 
\end{description}

\citet{2016Natur.529..181V} used Kepler photometry to measure the rotation
periods of 21 stars with precise ages derived using
asteroseismology. They find that some stars in this sample rotate significantly
faster than predicted by gyrochronology. van~Saders  et~al interpret these
observations as evidence for weakened magnetic braking in  stars more evolved
than the Sun. \citet{2016AN....337..810B} argue that this conclusion is a
consequence of including evolved stars and metal-poor stars in the sample.
Excluding these stars and one other from the sample, Barnes et~al find that
gyrochronological ages estimated using equation (\ref{gyro-eqn}) are
consistent with the age estimate from asteroseismology for the remaining eight
stars to within about 12\%. This is similar to the conclusion of
\citet{2014ApJ...790L..23D} who applied equation (\ref{gyro-eqn}) to 8
solar-type dwarf stars observed with Kepler with ages estimated using
asteroseismology of 1\,--\,8\,Gyr, although the uncertainties on the age
estimates in that case were quite large ($\approx 2$\,Gyr).

 There have been other claims that gyrochronological ages may be unreliable
for field stars \citep{2015MNRAS.450.1787A, 2017ApJ...835...75J,
2015A&A...581A...2K}. All of these studies suffer from contamination of the
samples studied by stars for which gyrochronology is not expected to produce
reliable results and, like all surveys,  will have some selection effects and
biases. This is currently an active topic of debate within the scientific
literature so it remains to be seen whether these issues are sufficient to
explain these apparent discrepancies.

\section{Observed trends in planet host star rotation}

 The three main sources of exoplanet discoveries that provide sufficient
numbers for statistical analysis of the host star properties are wide area
transit surveys, radial velocity surveys and the Kepler mission. We will look
at the rotation properties of the planet host stars from each type of survey
together since the selection effects and detection biases are broadly similar
within each of these groups.

\subsection{Radial velocity surveys}
 Surveys such as the Anglo-Australian planet search
\citep{2001ApJ...551..507T}, the \hbindex{HARPS} search for southern
extrasolar planets \citep{2010A&A...512A..48L} and the California Planet
Survey \citep{2010ApJ...721.1467H} typically target bright G- and K-type stars
that are pre-selected to have low rotation velocities and/or low levels of
magnetic activity. This maximises the sensitivity of these surveys to
long-period, low-mass planets, but makes it very difficult to draw conclusions
about the general rotational behaviour of stars with planets from these
surveys. Another complication is that the planets discovered by these surveys
rarely turn out to be transiting systems. There is usually very little
information available on the properties of non-transiting planets discovered in
radial velocity surveys beyond a lower limit to their mass, their orbital
periods, and the eccentricity of their orbits. 

 Comparisons between stars with and without detected planets from radial
velocity surveys find that any difference in \vsini\ between these groups is
either small or not significant
\citep{2011MNRAS.416L..80G, 2010MNRAS.408.1770A}.
 Apart from the selection effects and biases in the surveys, the rotation
periods of solar type stars are strongly correlated with their mass, age and
metallicity. This makes it very difficult to interpret any small difference in
\vsini\ between stars with and without detected planets,
particularly if these two groups come from samples with different selection
effects. For example, a small difference in \vsini\ could be due to the
stars with planets being older on average than stars without detected planets,
rather than due to the presence of a planet itself. The  convergence of
rotation periods due to magnetic braking also means that  any initial
difference in rotation periods between stars with and without planets will be
severely reduced over the lifetime of these stars except for the small number
of hot Jupiter systems found by these surveys.

\begin{figure}
\includegraphics[scale=.30]{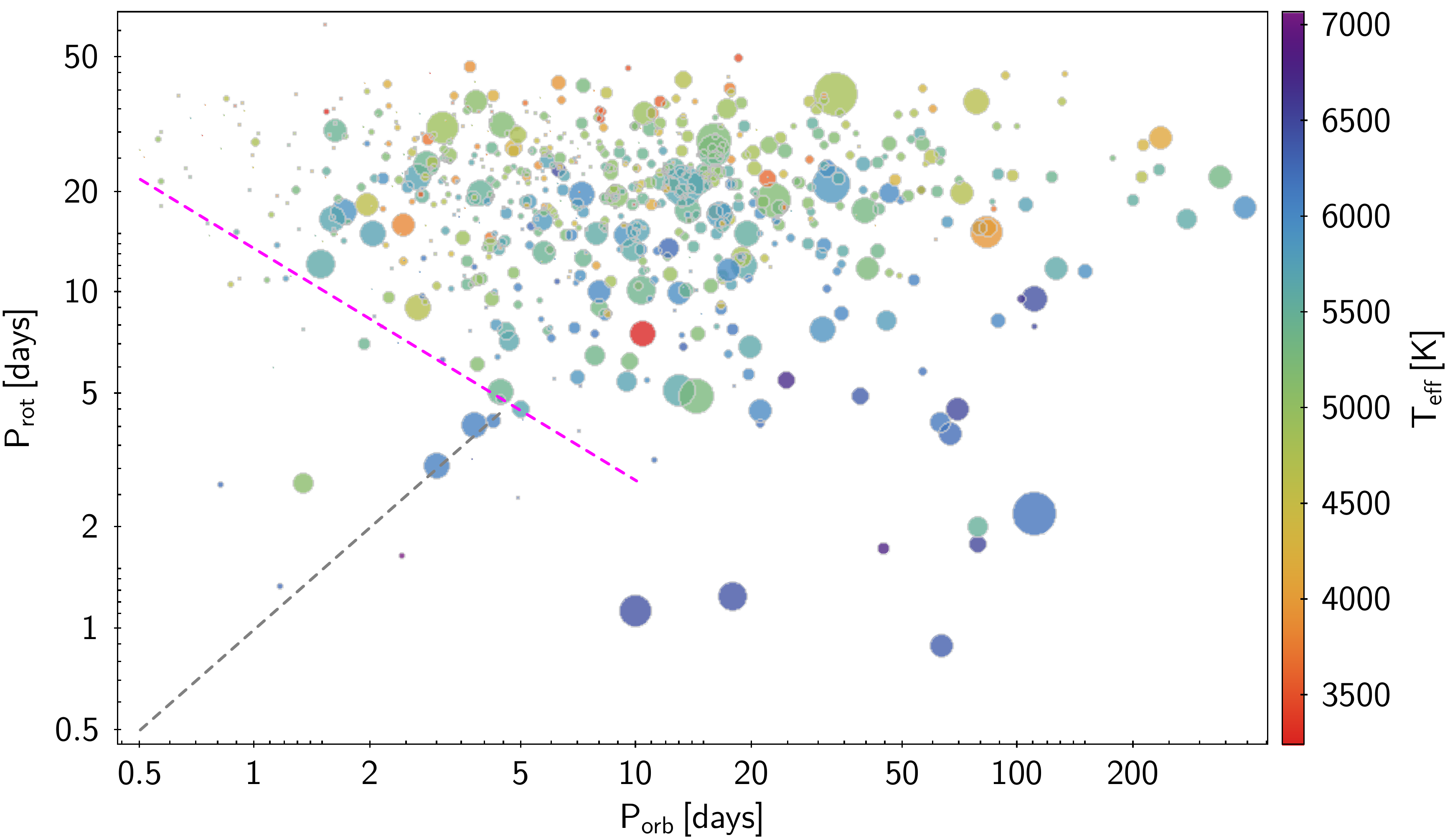}
\caption{Stellar rotation as a function of orbital period of the innermost
planet for the 737 Kepler object interest (KOIs) from Table 1 of
\citet{2013ApJ...775L..11M}. Symbol size is proportional to the logarithm of
the planet size. Note the lack of KOIs between the magenta dashed line and the
grey dashed line that indicates $P_{\rm rot} = P_{\rm orb}$.}
\label{fig:4}       
\end{figure}

\subsection{Kepler objects of interest}
 
 The Kepler mission obtained high-precision photometry for almost 200,000
stars during a four year period in a 115-square-degree region of the sky with
the aim of detecting transiting Earth-sized planets orbiting in the
\hbindex{habitable zone} of solar-like stars \citep{2008IAUS..249...17B,
2016AJ....152..158T}. Stars with Kepler data that are consistent with
planetary transits are designated as Kepler objects of interest (KOIs).
\citet{2013ApJ...766...81F} used Kepler data from the first 18 months of the
mission to show that approximately half of all stars observed by Kepler have
at least one planet in the period range 0.8\,--\,85\,days. This estimate is
based on 2,222 KOIs in a sample of 156,453 stars and accounts for the fact
that only about 1/40 planetary systems in this period range will show
detectable transits. 

 \citet{2013ApJ...775L..11M} measured the rotation periods for 737 KOI 
using Kepler photometry for 1919 KOIs with main-sequence host stars. Their
results are shown as a function of orbital period in Fig.~\ref{fig:4}. There
is a noticeable dearth of planets with short orbital periods among
rapidly-rotating stars. \citet{2014ApJ...786..139T} have interpreted these
observations as evidence for the destruction of planets in short-period orbits
in these systems, resulting in the spin-up of the host star. In this scenario,
the planet responsible for the transit in these KOIs is the survivor of what
was originally (and may still be) a multi-planet system. This scenario also
makes the testable prediction that systems close to the limit of destruction
(marked with a magenta line in Fig.~4) should, on average, be younger than
systems with comparable rotation periods but longer orbital periods. These
systems are currently in the process of angular momentum being transferred
from the planet's orbit to the star, so are not suitable targets for
gyrochronology. Indeed, \citet{2013MNRAS.436.1883W} find some evidence that
some stars with large planetary companions rotate synchronously for orbital
periods $P \mbox{\raisebox{-.3em}{$\stackrel{<}{\sim}$}}  10$ days, although
this claim has to be treated with some caution because there is a high rate of
false-positive planet detections in this part of the $P_{\rm
rot}$\,--\,$P_{\rm orb}$ diagram due to contamination of the Kepler photometry
by \hbindex{eclipsing binary stars}.

 \citet{2013MNRAS.436.1883W} also used published  \vsini\ measurements for a
small subset of their sample together with estimates of the stellar radius to
estimate $i_{\star}$ for these stars. They find that this method fails for
rotation periods $P \mbox{\raisebox{-.3em}{$\stackrel{>}{\sim}$}}  25$ days
because the rotational line broadening is too small to be measured reliably.
Among the shorter period systems they find most stars are consistent with
$i_{\star}\approx 90^{\circ}$ if some allowance is made for systematic errors
in the estimates of \vsini, as is expected if most transiting planets have
orbits with low obliquity. One exception is Kepler-9, a star that shows
transits from at least 3 planetary companions for which $i_{\star}\approx
45^{\circ}$. \citet{2014ApJ...783....9H} found similar results from their
sample of 32 KOIs for which they measured \vsini, and identified three other
\hbindex{multi-planet systems} that may have significant non-zero obliquity.
These systems are useful for studying the dynamical effects that can occur
during planet formation \citep{2014MNRAS.440.3532L, 2017MNRAS.tmp..186H}.

 The large number of stars that have planets that are not detected make it
difficult to interpret any observed difference in the rotation periods between
KOIs and non-KOIs in the Kepler sample, particularly given that there is a
detection bias against planets that transit rapidly rotating solar-type stars
because of additional noise from  stellar activity. 

\subsection{Wide area surveys and hot Jupiters}
  KOIs are typically quite faint, with most having apparent magnitudes in the
Kepler bandpass K$_{\rm p}=12$\,--\,16,  so most planet host stars bright
enough for detailed characterisation come from wide-area surveys such as WASP
\citep{2006PASP..118.1407P} and HAT \citep{2004PASP..116..266B}. The quality
of the photometry achievable by these ground-based surveys restricts them to
discovering planets with radii   $R
\mbox{\raisebox{-.3em}{$\stackrel{>}{\sim}$}}  0.5\,R_{\rm Jup}$ with orbital
periods less $P \mbox{\raisebox{-.3em}{$\stackrel{<}{\sim}$}}  10$ days, i.e.,
hot Jupiter systems. We also discuss in this sub-section results for other
bright hot-Jupiter systems such as those discovered by the CoRoT mission. 

\citet{2009MNRAS.396.1789P} found tentative evidence that the host stars of
some hot Jupiters rotate faster than typical stars of the same spectral type,
particularly for systems where the tidal interaction between the star and the
planet is large. This study was based on a sample of about 25 transiting
systems with published \vsini\ or $P_{\rm rot}$ estimates available at that
time. \cite{2015A&A...577A..90M} used Bayesian techniques to compare the age
estimates from stellar model isochrones to gyrochronological age estimates for
28 transiting exoplanet host stars with accurate mass and radius estimates and 
directly measured rotation periods. The gyrochronological age estimate  was
significantly lower than the isochronal age estimate for about half of the
stars in that sample. They found no clear correlation between the
gyrochronological age estimate  and the strength of the tidal force on the
star due to the innermost planet, leading them to conclude that tidal spin-up
is a reasonable explanation for this discrepancy in some cases, but not all.
For example, they  could find no satisfactory explanation for the discrepancy
between the young age for CoRoT-2 estimated from gyrochronology and supported
by its high \hbindex{lithium abundance}, and the extremely old age for its
K-type stellar companion inferred from its lack of magnetic activity. Maxted
et~al suggest that this may point to problems with the stellar models for
some planet host stars. 


\bibliographystyle{spbasicHBexo}  
\bibliography{maxted} 

\end{document}